\documentclass[alpha-refs]{wiley-article}

\usepackage{listings}
\usepackage{subcaption}
\usepackage{xcolor}
\definecolor{dkgreen}{rgb}{0,0.6,0}
\definecolor{gray}{rgb}{0.5,0.5,0.5}
\definecolor{mauve}{rgb}{0.58,0,0.82}

\usepackage[british]{babel}
\usepackage{upquote}
\usepackage{booktabs}
\usepackage{comment}

\usepackage{soul}
\usepackage{float}

\usepackage{graphicx}
\usepackage[space]{grffile}
\usepackage{latexsym}
\usepackage{textcomp}
\usepackage{longtable}
\usepackage{tabulary}
\usepackage{booktabs,array,multirow}
\usepackage{amsfonts,amsmath,amssymb}
\usepackage{url}
\usepackage{hyperref}
\hypersetup{colorlinks=true,pdfborder={0 0 0}}  
\usepackage{etoolbox}
\usepackage[sort&compress,capitalize]{cleveref}

\newif\iflatexml\latexmlfalse

\AtBeginDocument{\DeclareGraphicsExtensions{.pdf,.PDF,.eps,.EPS,.png,.PNG,.tif,.TIF,.jpg,.JPG,.jpeg,.JPEG}}

\usepackage[utf8]{inputenc}

\iflatexml

\else

\paperfield{Data science}
\corraddress{Fengnan Gao, School of Data Science, Fudan University, China}
\corremail{fngao@fudan.edu.cn}
\fundinginfo{National Natural Science Foundation of China, Grant Numbers: 11701095 and 11690013}

\fi

\papertype{Original Article}

\title{Pitfalls of amateur regression: The Dutch New Herring controversies} 

\author[1]{Gao, Fengnan}

\affil[1]{School of Data Science, Fudan University, China, and
              Shanghai Center for Mathematical Sciences}

\author[2]{Gill, Richard D.}

\affil[2]{Mathematical Institute, Leiden University, Netherlands}

\runningauthor{Gao and Gill}

\begin{document}

\maketitle
\begin{abstract}

Applying simple linear regression models, an economist analysed a published dataset from an influential annual ranking in 2016 and 2017 of consumer outlets for Dutch New Herring and concluded that the ranking was manipulated.  His finding was promoted by his university in national and international media, and this led to public outrage and ensuing discontinuation of the survey.  We reconstitute the dataset, correcting errors and exposing features already important in a descriptive analysis of the data.  The economist has continued his investigations, and in a follow-up publication repeats the same accusations.  We point out errors in his reasoning and show that alleged evidence for deliberate manipulation of the ranking could easily be an artefact of specification errors.  Temporal and spatial factors are both important and complex, and their effects cannot be captured using simple models, given the small sample sizes and many factors determining perceived taste of a food product. 


\end{abstract}



\section{Introduction}
\label{intro}


This paper presents a case-study of a problem in which simple regression analyses were used to make suggestions of causality, in a context with both spatial and temporal aspects. The findings were well publicized, and this badly damaged the interests of several commercial concerns as well as individual persons. Our study illustrates the pernicious effects of the eagerness of university PR departments to promote research results of societal interest, even if tentative and ``unripe''. Nowadays, anyone can perform standard statistical analysis without understanding of the conditions under which they could be valid, certainly if causal conclusions are desired. The damage caused by such activities is hard to correct; simply asserting that correlation does not prove causality does not convince anyone. Careful research documentation of sound counterarguments is vital, and the present paper does just that. We moreover attempt to make the case, the questions, and the data accessible to theoretical statisticians and hope that some will come up with interesting alternative analyses.  A main aim is to underscore the synergy which is absolutely required in applied ``data science'' of subject matter knowledge, theoretical statistical understanding, and modern computational hardware/software. 

In this introductory section, we first briefly describe the case, and then give an outline of the rest of the paper.

For many years, a popular Rotterdam based newspaper \emph{Algemeen Dagblad} (AD), published an immensely influential annual survey of a typically Dutch seasonal product: Dutch New Herring (\emph{Hollandse Nieuwe}). This data included not only a ranking of all participating outlets and their final scores but also numerical, qualitative, and verbal evaluations of many features of the product being offered. A position in the top ten was highly coveted. Being in the bottom ten was a disaster. The verbal evaluations were often pithy.

However, rumours circulated that the test was biased.  Every year, the herring test was performed by the same team of three tasters, whose leader was consultant to a wholesale company called \emph{Atlantic} based in Scheveningen, not far from Rotterdam (both cities on the West Coast of the Netherlands). He offered a popular course on herring preparation.  His career was  dedicated to promotion of ``Dutch New Herring'', and he had earlier successfully managed to obtain the European Union (EU) legal protection for this designation. 

Enter economist Dr Ben Vollaard of Tilburg University. Himself partial to a tasty Dutch New Herring, he learnt in 2017 from his local fishmonger about complaints then circulating about the AD Herring Test. Tilburg is somewhat inland. Consumers in different regions of the country have probably developed different tastes in Dutch New Herring, and a common complaint was that the AD herring testers had a Rotterdam bias.
Vollaard downloaded the data published on AD's website on $144$ participating outlets in 2016, and $148$ in 2017, and ran a linear regression analysis (with a $292 \times 21$ design matrix), attempting to predict the published final score for each outlet in each year, using as explanatory variables the testing team’s evaluations of the herring according to twelve criteria of various nature: subjectively judged features such as ripeness and cleaning; numerical variables such as weight, price, temperature; laboratory measurements of fat content and microbiological contamination. Most of the numerical variables were modelled by using dummy variables after discretization into a few categories, and some categorical variables had some categories grouped. A single indicator variable for ``distance from Rotterdam’’ (greater than 30 kilometres) was used to test for regional bias.

It had a just significant negative effect, lowering the final score by about $0.4$. Given the supreme importance of getting the highest possible score, 10, a loss of half a point could make a huge difference to a new outlet going all out for a top score and hence position in the ``top ten'' of the resulting ranking. 
Vollaard concluded in a working paper \citet{vollaard1} ``\emph{everything indicates that herring sales points in Rotterdam and the surrounding area receive a higher score in the AD Herring Test than can be explained from the quality of the herring served}''. His university put out a press release which drew a lot of media attention. 
A second working paper \citet{vollaard2} focussed on the conflict of interest concerning wholesale outlet \emph{Atlantic}. By contacting outlets directly, Vollaard  identified $20$ outlets in the sample whose herring he thought have been supplied by that company.  As was already known, \emph{Atlantic}-supplied herring outlets tended to have good final scores, and a few of them were regularly in the top ten. 
The author did not report the fact that the dummy variable for being supplied by \emph{Atlantic} was not statistically significant when added to the model he had already developed. Instead, he came up with a rather different argument from the one which he had used for the Rotterdam bias question. He argued that his regression model showed that the Atlantic outlets were being given an \emph{almost 2-point advantage} based on subjectively scored characteristics.  Another university press release led to more media attention. The work was reported in \emph{The Economist} \citep{vollaardEconomist2017}. The AD suspended its herring test but fought back with legal action against Vollaard through a scientific integrity complaint. 

Vollaard was judged not guilty of any violation of scientific integrity, but short-comings of his research were confirmed and further research was deemed necessary. The key variable ``Atlantic supplied'' had important errors. He continued his investigations, joined by a former colleague, and recently published \citet{vollaard2021bias} with the same accusation of favouritism toward Atlantic supplied outlets, but yet again quite different arguments for them. 
Some but not all of the statistical shortcomings of the original two working papers are addressed, and some interesting new ideas are brought into the analysis, as well as an attempt to incorporate the verbal assessments of ``taste'' into the analysis. 
However, our main statistical issues remain prominent in the new paper. 

The present paper analyses the same data with a view to understanding whether the claim of effective and serious favouritism can be given empirical support from the data. This is a case where society is asking causal questions, yet the data is clearly a self-selecting sample, the ``treatment'' (supplier = \emph{Atlantic}) is not randomized or blinded. There is every reason that any particular linear regression model specifying the effect of twelve measured explanatory variables \emph{must be wrong}, but could it still be \emph{useful}? 

There are major temporal and spatial issues. 
The AD herring test started in the Rotterdam region but slowly expanded to the whole country.  Just a small proportion of last year's participants enter themselves again and moreover AD did its best to have last year's top ten tested again. There is a major problem of confounding of the effects of space and time and ``Atlantic-supplied'', with new entrants to the AD Herring Test tending to come from more distant locations and often doing poorly on a first attempt.  Clearly, drawing causal conclusions from such a small and self-recruiting sample is fraught with danger. We will treat the statistical analyses both of Vollaard and (new ones) of our own as exploratory data analysis, as descriptive tools.  Even if (for instance) a particular linear regression model cannot be seriously treated as a causal model and the ``sample'' is not a random sample from a well-defined population, we believe that statistical significance still has a role to play in that context at the very least as ``sensitivity analysis''.

Theoretical developments of jackknife and bootstrap methods --- both popular methods in sensitivity analysis --- suggest that these resampling techniques can produce similar inferences to classical methodology based on parametric model assumptions. Conversely, this means that parametric inferences correspond to purely data analytic sensitivity analysis. For example, if a coefficient in a regression model is not significant at the 5\% level based on the conventional normal theory-based t-statistic, it is likely that 95\% of bootstrap confidence intervals contain the value 0.

It's worth noting that we only interpret regression coefficients descriptively rather than causally. The ``effect'' of a variable (supplier = Atlantic) is merely the difference in the average score between Atlantic-supplied outlets and other outlets similar with respect to the other covariates in the model, and this requires assuming that the effects of various factors are additive. However, this assumption may be incorrect, especially in light of the fact that linearity assumptions regarding the effect of numerical covariates are also additivity assumptions.

Particularly, the statistical significances in Vollaard's papers, including in recent \citet{vollaard2021bias}, suffer from the major problem of instability of estimates under minor changes in specification of the variables, and to errors in the data.
If statistical significance cannot be established based on standard distributional assumptions for a certain factor, then data-analytic measures of numerical stability under small changes to the dataset will show corresponding instability or sensitivity. Yet such small changes could be inherent to the dataset, like those due to errors in measurement of some variables or small changes in the specification of the effects of some variables.

A big danger in exploratory analyses is cherry-picking, especially if researchers have a strong desire to find support for a certain causal hypothesis. This clearly applies to both ``teams'' (the present authors versus Vollaard and van Ours); for our conflict of interest, see \cref{COI}. Certainly, the whole concept of the AD Herring Test was blemished by the \emph{existence} of a conflict of interest. One of the three herring tasters gave courses on the right way to prepare \emph{Hollandse Nieuwe} and on how it should taste, and was consultant to one wholesaler. He was effectively evaluating his own students. He was the acknowledged expert on \emph{Hollandse Nieuwe} in the team of three.  But Vollaard and van Ours want their statistical analyses to support the strong and damaging conclusion that the AD's final ranking was seriously affected by favouritism.

In the meantime, the AD Herring Test has been rebooted by another newspaper. There is now in principle seven years more data from a survey specifically designed to avoid the possibility of any favouritism. There also exists more than 30 years of data from past surveys. We hope that the present ``cautionary tale'' will stimulate discussion, leading to new analyses, possibly on new data, by unbiased scientists. The data of 2016 and 2017, and our analysis scripts, are available on our GitHub repository,\footnote{\url{https://github.com/gaofengnan/dutch-new-herring}} and we would love to see new analyses with new tools, and especially new analysis of new data.

The paper is organized as follows. In the next \cref{dutchnewherring} we provide further details about what is special about Dutch New Herring, since the ``data science'' which will follow needs to be informed by relevant subject matter knowledge. We then, in \cref{adherringtest}, briefly describe how the AD Herring Test worked. Then follows, in \cref{vollaardPaper1Model2}, the main analysis of Vollaard's first working paper \citet{vollaard1}. This enables us to discuss some key features of the dataset which, we argue, must be taken account of. After that, in \cref{atlantic} we go into the issue of possible favouritism toward the outlets supplied by wholesaler \emph{Atlantic}, and explored in the second working paper \citet{vollaard2} and in the finally published paper  \citet{vollaard2021bias}. In particular, we apply a technique from the theory of causality for investigating bias; essentially it is a nonparametric estimate of the effect of interest (using the words ``effect of'' in the statistical sense of ``difference associated with''). 
We also take a more refined look at the spatial and temporal features in the data and argue that the question of bias favouring Atlantic outlets is confounded with both. 
There is a tendency for the new entrants to come from locations more distant from the coast and in regions where Dutch new herring is consumed less. \emph{Atlantic} too has been extending its operations. 
Finally, the factors determining the taste of Dutch New Herring, including spatial factors, are too complex and too interrelated for their effects to be separated with what is a rather small dataset, with a glimpse at just two time points of an evolving geographical phenomenon. 
\Cref{whatnext} discusses a new Herring Test based on another city, Leiden.  The post-AD test also had national aspirations, and attempted to correct some obviously flawed features of the classic test. However, it seemed that it did not succeed in retaining its popular appeal.

In \cref{conclusion} we summarize our findings. We conclude that there is nothing in the data to suggest that the testing team abused their conflict of interest. 

\section{What makes herring Dutch New Herring}\label{dutchnewherring}

Every nation around the North Sea has traditional ways of preparing North Atlantic herring. For centuries, herring has been a staple diet of the masses. It is typically caught when the North Atlantic herring population comes together at its spawning grounds, one of them being in the \foreignlanguage{Norwegian}{Skagerrak}, between Norway and Denmark. Just once a year there is an opportunity for fishers to catch enormous quantities of a particular extremely nutritious fish, at the height of their physical condition, about to engage in an orgy of procreation. 

Traditionally, the Dutch herring fleet brought in the first of the new herring catch mid-June. The separate barrels in the very first catch are auctioned and a huge price (given to charity) is paid for the very first barrel. Very soon, fishmongers, from big companies with a chain of stores and restaurants, to supermarket chains, to small businesses selling fish in local shops and street markets are offering Dutch New Herring to their customers. It's a traditional delicacy, and nowadays, thanks to refrigeration, it can be sold the whole year long, though it may only be called \emph{new} herring for the first few months. Nowadays, the fish arrives in refrigerated lorries from Denmark, no longer in Dutch fishing boats at Scheveningen harbour. Moreover, for reasons of public health (killing off possible parasites) the fish must at some point have been frozen at a sufficiently low temperature for a sufficiently long time period. One could argue that traditional preservation methods are superfluous. But, they do have distinctive and treasured gastronomical consequences, and hence are treasured by consumers and generate economic opportunities for businesses. 

What makes a Dutch new herring any different from the herring brought to other North Sea and Baltic Sea harbours? The organs of the fish should be removed when they were caught, and the fish kept in lightly salted water. But two internal organs are left, a fish's equivalent to our pancreas and kidney. The fish's pancreas contains enzymes which slowly transform some protein into fat and this process is responsible for a special almost creamy taste which is much treasured by Dutch consumers, as well as those in neighbouring countries. 


\section{The AD Herring Test}\label{adherringtest}

For many years, the Rotterdam based newspaper \emph{Algemene Dagblad} (AD) carried out an annual comparison of the quality of the product offered in a sample of consumer outlets. A fixed team of three (a long time professional herring expert, a senior editor of AD, and a younger journalist) paid surprise visits to the typical small fishmonger's shops and market stalls where customers can order portions of fish and eat them on the premises, or even just standing in a busy food market or shopping street. 
The team evaluated how well the fish has been prepared, preferring especially that the fish have not been cleaned in advance but that they are carefully and properly prepared in front of the client. They observed adherence to basic hygiene rules. They judged the taste and checked the temperature at which it is given to the customer: by law it may not be above 7 degrees, though some latitude in this is tolerated by food inspectors. And they recorded the price. An important though obviously somewhat subjective characteristic is ``ripeness''. Expert tasters distinguish Dutch new herring which has not ripened (matured) at all: it is designated ``green''. After that comes lightly, well, or too much matured, and after that, rotten. The ripening process is of chemical nature and is due to time and temperature: fat becomes oil and oil becomes rancid. 

At the conclusion of their visit they agreed together on a ``provisional overall score'' as well as classifications of ripeness and of how well the herring was cleaned.  The provisional scores range from 0 to 10, where 10 is perfection;  below 5.5 is a failing grade. The intended interpretation of a particular score follows Dutch traditions in education from primary school through to university, and we will spend some more words later on some fine details of the scoring.  

\begin{figure}[htp]
\includegraphics[width=0.9\textwidth]{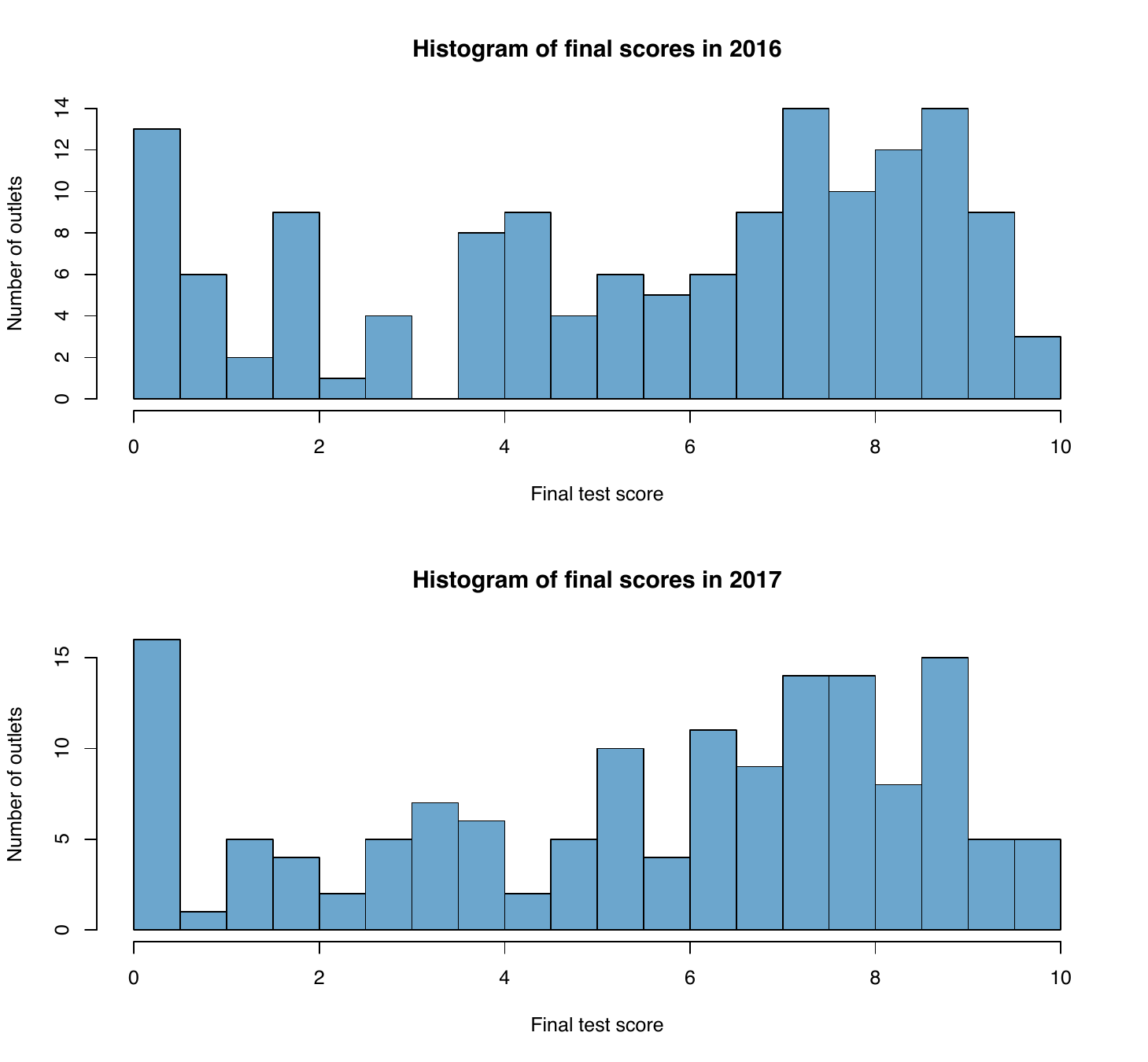}
\caption{Histogram of the final test scores 2016 and 2017, respectively, $N=144$ for 2016 and $N=148$ for 2016.}
\label{fig:1}       
\end{figure}

Fish was also sent to a lab for a number of measurements: weight per fish, fat percentage, signs of microbiological contamination. On reception of the results, the team produced a final overall score. Outlets which sold fish that was definitely rotten, definitely contaminated with harmful bacteria, or definitely too warm got a zero grade. The outlets were ranked and the ten highest ranking outlets were visited again, the previous evaluations checked, and scores adjusted in order to break ties.  The final ranking was published in the newspaper, and put in its entirety on internet, together with the separate evaluations mentioned so far. 
One sees from the histogram \cref{fig:1}, that in both 2016 and 2017, more than 40\% of the outlets got a failing grade; almost 10\% were essentially disqualified, by being given a grade of zero. The distribution looks nicely smooth except for the peak at zero, which really means that their wares did not satisfy minimal legal health requirements or were considered too disgusting to touch. The verbal judgements on the website explained such decisions, often in pretty sharp terms.

\section{The analysis which started it all}\label{vollaardPaper1Model2}

Here is the main result of  \citet{vollaard1}; it is the second of two models presented there. The first model simply did not include the variable ``distance from Rotterdam''.

{
\begin{lstlisting}
lm(formula = finalscore ~ 
                    weight + temp + price + fat + fresh + micro + 
                    ripeness + cleaning + yr2017 + k30)
 
Residuals:
     Min      1Q  Median      3Q     Max 
 -4.0611 -0.5993  0.0552  0.8095  3.9866

Residual standard error: 1.282 on 274 degrees of freedom
Multiple R-squared:  0.8268, Adjusted R-squared:  0.816 
F-statistic: 76.92 on 17 and 274 DF,  p-value: < 2.2e-16


Coefficients:
                  Estimate   Std.Error  t-value  Pr(>|t|)
                              
Intercept         4.139005   0.727812   5.687 3.31e-08 ***

weight (grams)    0.039137   0.009726   4.024 7.41e-05 ***
 
temp
    < 7 deg       reference category
    7--10 deg    -0.685962   0.193448  -3.546 0.000460 ***
    > 10 deg     -1.793139   0.223113  -8.037 2.77e-14 ***
 
fat
    < 10%         reference category
    10--14%       0.172845   0.197387   0.876 0.381978    
    > 14%         0.581602   0.250033   2.326 0.020743 *  
    
price per 100g, Euro
   < 2.84         reference category
   2.84--3.48     0.47639    0.21211    2.246 0.025509 *  
   > 3.48         0.27148    0.27406    0.991 0.322777   
 
freshly prepared  1.817081   0.200335   9.070 < 2e-16 ***
 
micro
    very good    reference category
    adequate     -0.161412   0.315593  -0.511 0.609443    
    bad          -0.618397   0.448309  -1.379 0.168897    
    warning      -0.151143   0.291129  -0.519 0.604067    
    reject       -2.279099   0.683553  -3.334 0.000973 ***
 
ripeness
    mild         reference category
    average      -0.377860   0.336139  -1.124 0.261947    
    strong       -1.930692   0.386549  -4.995 1.05e-06 ***
    rotten       -4.598752   0.503490  -9.134  < 2e-16 ***
 
cleaning
    very good    reference category
    good         -0.983911   0.210504  -4.674 4.64e-06 ***
    poor         -1.716668   0.223459  -7.682 2.79e-13 ***
    bad          -2.761112   0.439442  -6.283 1.30e-09 ***

year
    2016          reference category
    2017          0.208296   0.174740   1.192 0.234279

distance from Rotterdam
    < 30 km       reference category
    > 30 km       -0.37173   0.17278   -2.151 0.032322 *  
     
--
Signif. codes:  0 '***' 0.001 '**' 0.01 '*' 0.05 '.' 0.1 ' ' 1
\end{lstlisting}
}

The testing team prefers fatty and pricier herring, properly cool, mildly matured, freshly prepared on site, and well-cleaned on site too. We have a delightful amount of statistical significance. Especially significant for the author (a p-value of 3\%), is effect of distance from Rotterdam: outlets more than 30 km from the city lose one third of a point (recall, it is a ten-point scale). The effect of \texttt{year} is insignificant and small (in the second year, the scores are on average perhaps just a bit larger). The value of $R^2$, just above $80\%$, would be a delight to any micro-economist. The estimated standard deviation of the error term is however about $1.3$ which means that the model does not do well if one is interested in distinguishing grades familiar to those used throughout Dutch education system. For instance, $6$, $7$, $8$, $9$, $10$ have the verbal equivalents ``sufficient'', ``more than sufficient'', ``good'', ``very good'', ``excellent'', where ``sufficient'' means: enough to count as a ``pass''. This model does not predict well at all.

There are some curious features of Vollaard's chosen model: some numerical variables (\texttt{temp}, \texttt{fat}, and \texttt{price}) have been converted into categorical variables by some choice of just two cut points each, while \texttt{weight} is taken as numerical, with no explanation of the choice. One should worry about interactions and about additivity. Certainly one should worry about model fit.

\begin{figure}[htp]
\begin{subfigure}{0.5\textwidth}
\includegraphics[width=0.95\textwidth]{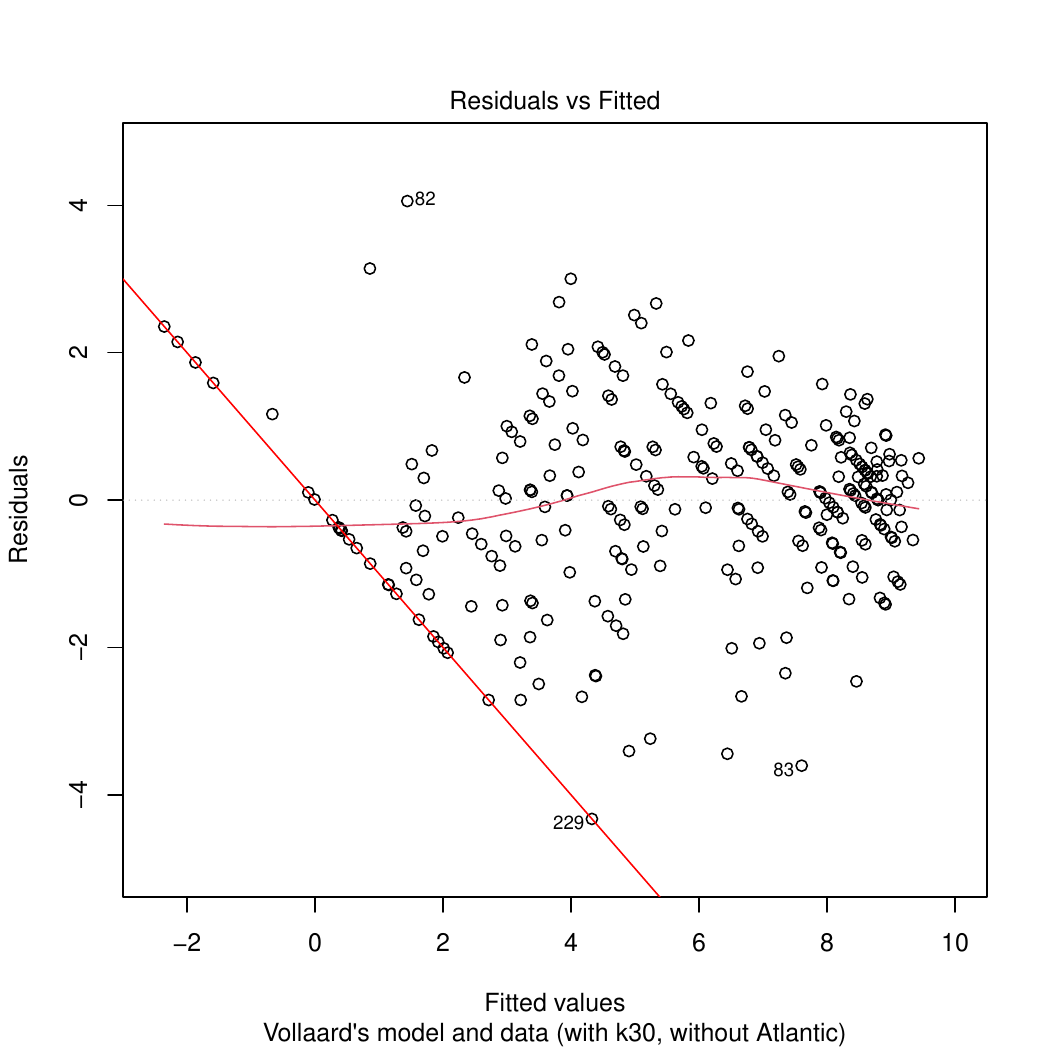}
\end{subfigure}
\begin{subfigure}{0.5\textwidth}
\includegraphics[width=0.95\textwidth]{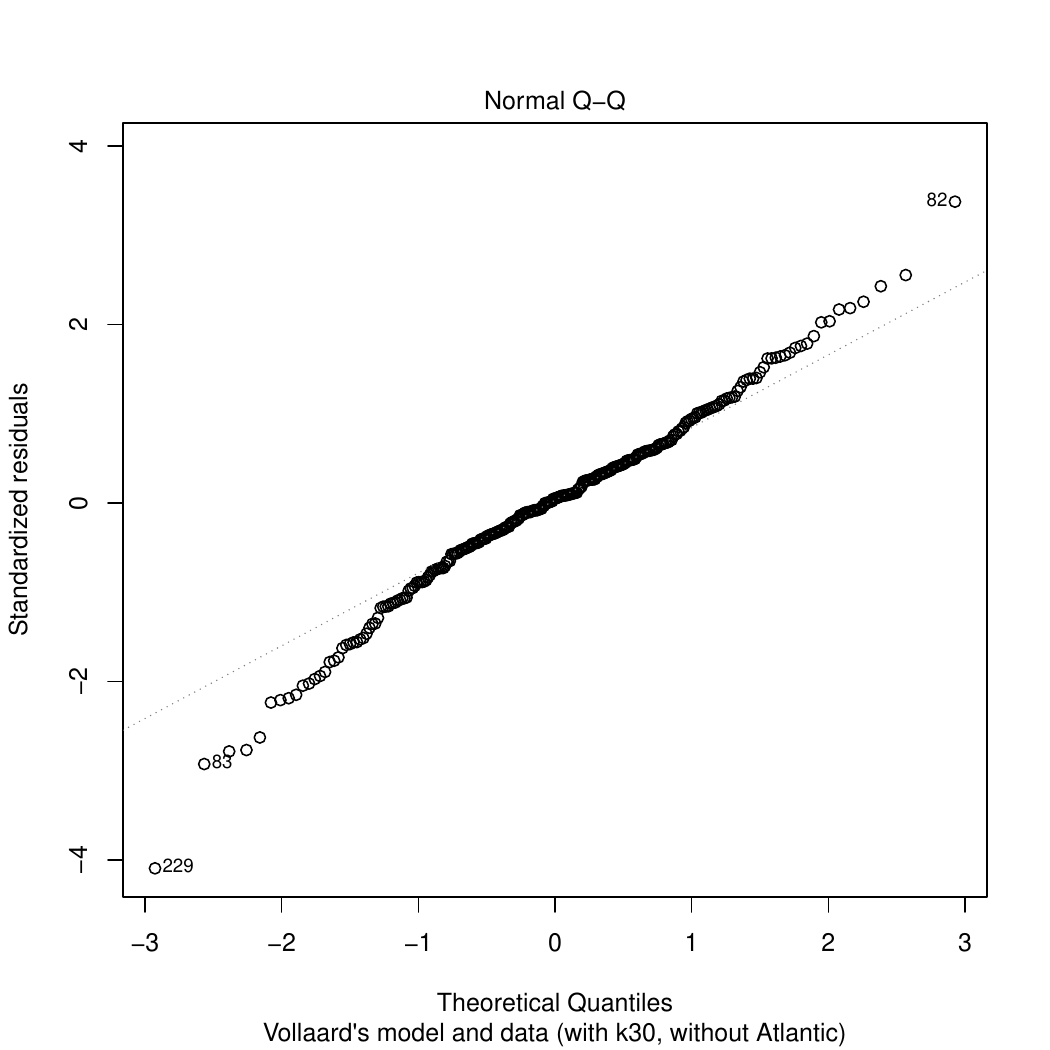}
\end{subfigure}
\\[5pt]
\begin{subfigure}{0.5\textwidth}
\includegraphics[width=0.95\textwidth]{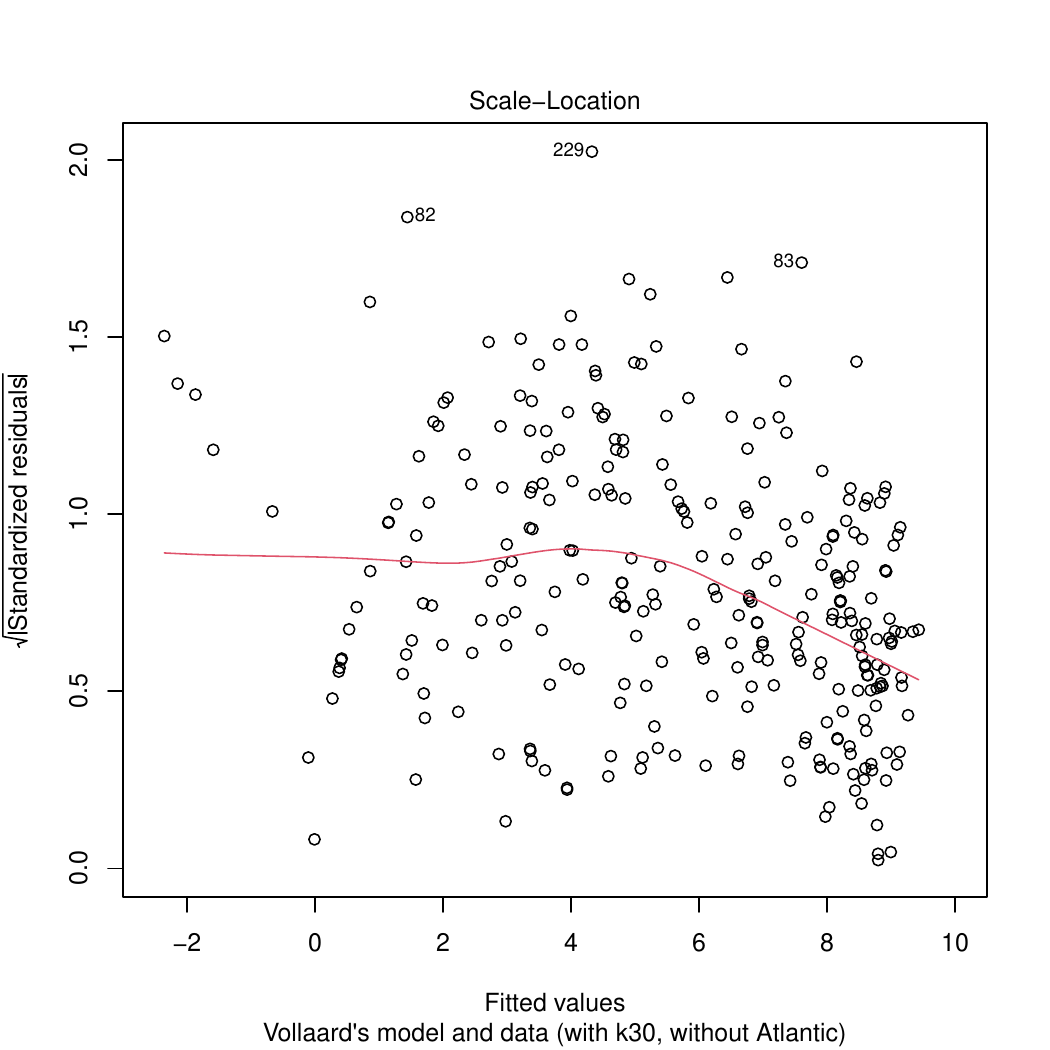}
\end{subfigure}
\begin{subfigure}{0.5\textwidth}
\includegraphics[width=0.95\textwidth]{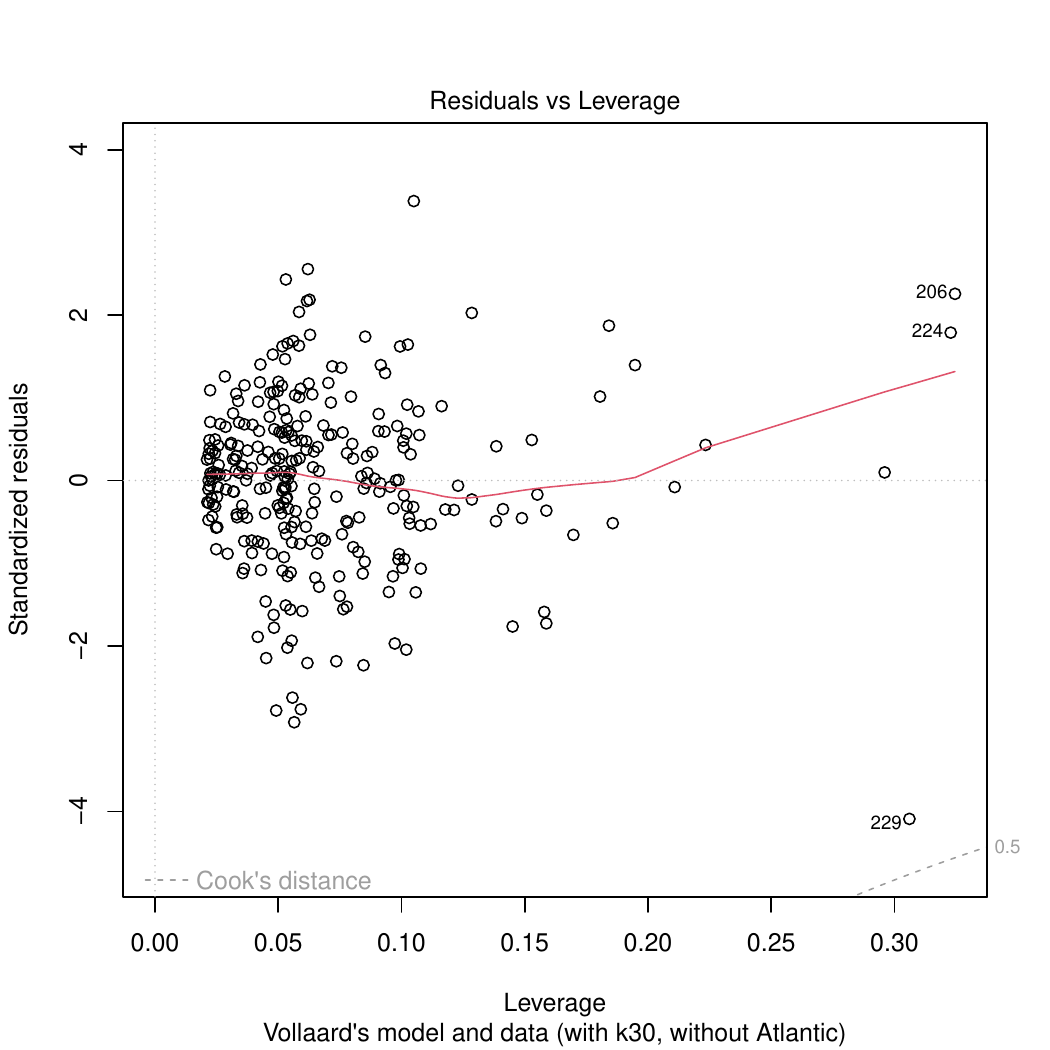}
\end{subfigure}
\caption{Diagnostic plots (residual analysis). We added the line  $y = -x$ to the first plot.}
\label{fig:model-vali}
\end{figure}

We add to the estimated regression model also \texttt{R}'s standard diagnostic plots in \cref{fig:model-vali}, to which we have made one addition, as well as changing the default limits to the x-axis in two of the plots. The dependent variable lies in the interval $[0, 10]$. There are no predicted values as large as 10, but plenty smaller than 0.

The plots do not look good. The error distribution has heavy tails on both sides and three observations are marked as possible outliers.  We see residuals as large as $\pm 4$ though the estimated standard deviation of the error terms is about $1.3$; two standard deviations is about $2.5$. There is a serious issue with the observations which got a final score of zero: notice the downward sloping straight line, lower envelope of the scatter plot, bottom left of the plot of residuals against fitted values. The observations on this line (the line $y = - x$) have observed score zero, residual equals the negative of the predicted value. There are predicted values smaller than $-2$. These are outlets which have essentially been disqualified on grounds of violation of basic hygiene laws; most of the explanatory variables were irrelevant. 
Recall that the panellists were in practice lenient in allowing for higher temperatures than the regulatory maximum of 7 degrees. 

The model gives all outlets which were given the score $10$ ("excellent") a negative residual. Because of the team's necessarily downward adjustment of final scores in order to break ties in the top 10, there can only be one such observation in each year. The reader should be able to spot those two instances in the first plot. Again here, we see that the linearity assumption of the model really makes no sense. 

\begin{figure}
\centering
\includegraphics[width=0.6\textwidth]{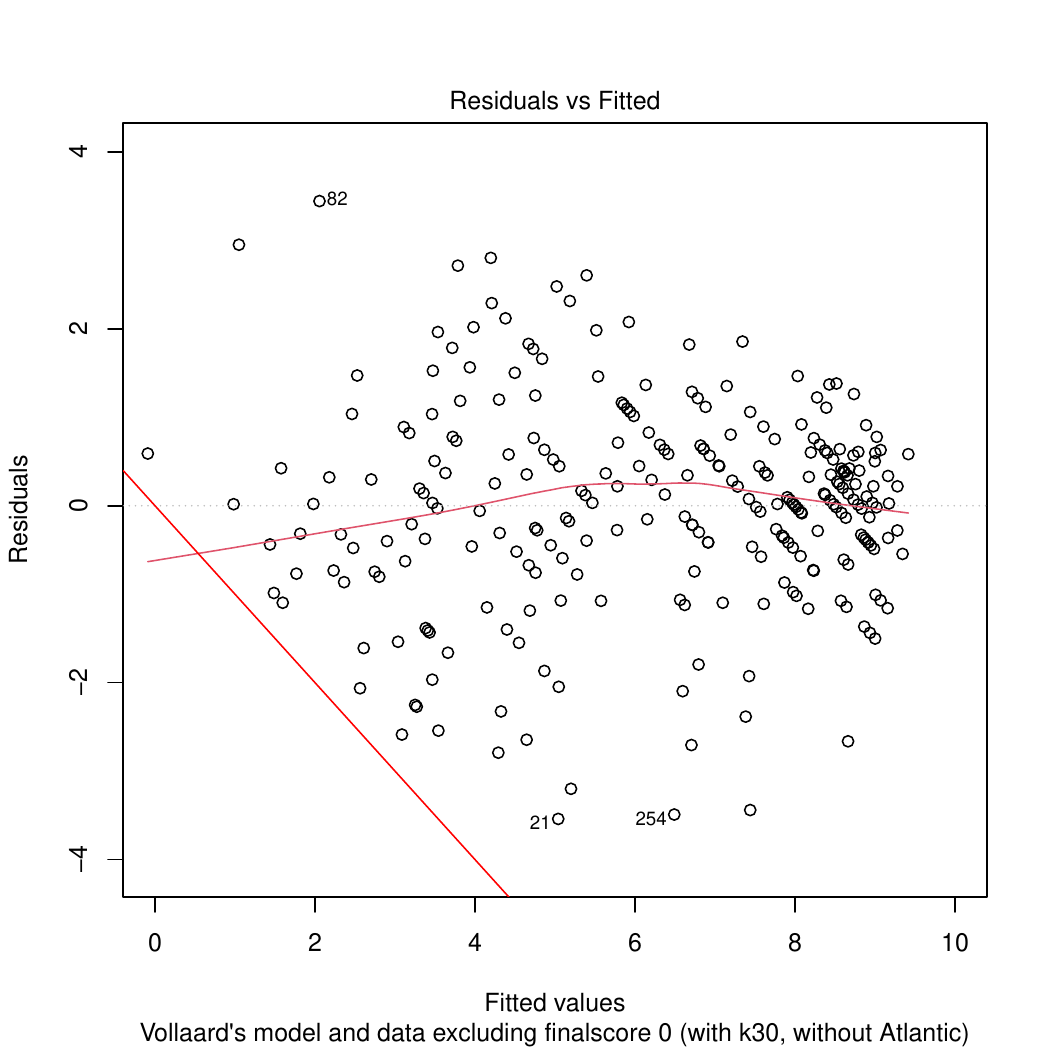}
\caption{Residuals against fitted values when ``disqualified'' outlets are removed. Spot the two years' winners.}
\label{fig:model-vali-subset}
\end{figure}

When we leave out the ``disqualified'' outlets, the residual diagnostic plots look much cleaner. The parameter estimates and their estimated standard errors are not much different.  Here we exhibit just the first diagnostic plot in \cref{fig:model-vali-subset}: the plot of residuals against fitted values.
The cloud of points has a clear shape which we easily could have predicted. Residuals can be large in either direction when the predicted value is 5 or 6, they cannot be large and negative when the predicted value is near zero, nor large and positive when the predicted value is close to 10. Apart from a few exceptionally large outliers, the scatter plot is lens-shaped, \raisebox{1pt}{$\frac\frown\smile$}: the variation is large in the middle and small at both ends. One might model this heteroscedasticity by supposing that the error variance in a model for final score divided by 10, thus expressed as a number $p$ in the unit interval $[0, 1]$, has a binomial type variance $c p (1 - p)$, or better $a + b p (1 - p)$. One can also be ambitious and measure the variance of the error as a smooth function of the predicted value. We tried out the standard LOESS method \citep*{cleveland1979robust}, 
which resulted in an estimate of the general shape just mentioned.  We then fit the model again using the LOESS estimate of variance to provide weights.  But not much changed as far as estimated effects and estimated standard errors were concerned.

An alternative way to describe the data follows from the fact that the herring tasters were essentially after a ranking of the participating outlets. The actual scores stand for ordered qualifications. The variable we are trying to explain is ordinal. Small differences among the high scores are important. Or one might even estimate non-parametrically a monotone transformation of the dependent variable, perhaps improving predictability by a simple linear function, and perhaps stabilizing the variance at the same time (think of Fisher's arc-sine transformation applied to score divided by 10). We tried out the standard ACE method \citep*{breiman1985estimating},
which came up with piecewise linear transformation with three very small jumps upwards and small changes of slope, at the values 8, 8.5 and 9. The scores above 9 were pulled together (a smaller slope). The analysis reflects the testers' tendency to heap scores at values with special meaning and their special procedure for breaking ties in the top ten, which spreads them out. It has next to no effect on the fitted linear regression model.

There is another issue that should have been addressed in Vollaard's working papers. We know that some observations come in pairs --- the same outlet evaluated in two subsequent years. AD tried to get each year's top 10 to come back for next year's test, and often they did.  We turned to the AD (and the original webpages of AD) to sort this out. After the outlets had been identified, we analysed the two years of data together, correcting for correlation between outlets appearing in two subsequent years. There were only 23 such pairs of observations. It turned out that the correlation between the residuals of the same outlet participating in two subsequent years was quite large, as one could have expected, about $0.7$. However, their number is fairly small, and this had little effect on the model estimates. Taking account of it slightly increases the standard errors of estimated coefficients.  Alternatively, correction for autocorrelations could easily be made superfluous by dropping all outlets appearing for the second year in succession. Then we would have two years of data, in the second year only of ``newly nominated'' outlets. Perhaps we should have made the same restriction to the first year, but that would require going back to older AD web pages.

Notice that dropping ``return'' outlets removes many of the top ten in the second of the two years and therefore removes several second-year Atlantic supplied outlets, which brings us to the topic of the next section.

\section{Was there favouritism toward Atlantic supplied outlets?}\label{atlantic}

\subsection{The first argument}

In \citet{vollaard2} the author turned to the question of favouritism specifically of Atlantic supplied outlets, and this was again the main theme of  \citet{vollaard2021bias}.  Atlantic had declined to inform him which outlets had served herring they had originally supplied.  He called up fishmonger after fishmonger and asked them whether the AD team had been served Atlantic herring. 

It is likely that Vollaard first investigated the possibility of bias by adding his variable ``AD supplied'' as a dummy variable to his model.  If so, he would have been disappointed, because had he done so, its effect would not have been significant, and anyway rather small, similar to the effect of distance from Rotterdam (as modelled by him).  However, he turned to a new argument for bias. Many of the explanatory variables in the model have mean values on his 20 presumed Atlantic outlets which lead to high predicted scores. He used his model to predict the mean final score of Atlantic outlets, and the mean final score of non-Atlantic outlets. Both of these predictions are, unsurprisingly, close to the corresponding observed averages (his estimated model ``fits'' his Atlantic outlets just fine). The difference is about 3.5 points. He then noted that this difference can be separated into two parts: the part due to the ``objective'' variables (weight, fat content, temperature, cleaned on site in view of the client, ...) and the part due to the ``subjective'' variables (especially: cleaning, ripeness). It turned out that the two parts were each responsible for about half of the just mentioned difference; which means a close to 2-point difference. 

By the way, the model had also been slightly modified. There is now an explanatory variable ``top ten'', which not surprisingly comes out highly significant.  We agree that the extra steps taken by the test team to sort out the top ten need to be taken account of, but using a feature of the final score to explain the final score makes no sense.

Vollaard concluded from this analysis that the testers' evaluation is dominated by subjective features of the served fish, and that this had given the Atlantic outlets their privileged position close to the top of the ranking.  He wrote that the Atlantic outlets had been given a two-point advantage due to favouritism. (He agreed that he had not \emph{proved} this, since correlation is not causation, but he did emphasize that this is what his analysis \emph{suggested}, and this was what he believed.)  

The argument is however very weak.  Whether bones and fins or scales are properly removed is hardly subjective.  Whether a herring is green, nicely matured, or gone rancid, is not so subjective, though fine points of grading of maturity can be matters of taste. Actual consumer preference for different degrees of ripening may well vary by region, but suggesting that the team deliberately used the finer shades of distinction to favour particular outlets is a serious accusation. Suggesting that it generates a two-point systematic advantage seems to us simply wrong and irresponsible.

Incidentally, AD claimed that Vollaard's classification of outlets supplied by \emph{Atlantic} was badly wrong. One \emph{Atlantic} outlet had received, in one year, a final score of 0.5, and that was obviously inconsistent with the average reported by Vollaard since his number of Atlantic outlets (9 in one year, 11 in the next) was so small that a score of 0.5 would have resulted in a lower average than the one he reported. AD supplied us with a list of Atlantic outlets obtained from Atlantic itself. The total number went up by 10 while one or two of Vollaard's Atlantic outlets were removed. There is a problematic issue here: possibly Atlantic sells different ``quality grades'' of herring at different prices (this is clearly a sensitive issue, which neither Atlantic nor outlet might like to reveal). Next, while Atlantic have easily identified which outlets they supplied in any year, there is no guarantee that they were the only wholesaler who supplied any particular fishmonger. So Atlantic could well be ignorant of whether a particular fishmonger served \emph{their} Dutch new herring on the day that the herring testers paid their visit. Hopefully the fishmonger does know, but will they tell?

Here are some further discoveries while we were reconstituting the original dataset.  Vollaard had been obliged to make adjustments to the published final scores.  In both years there were scores registered such as $8-$ or $8+$, meant to indicate ``nearly an 8'' or ``a really good 8'' respectively, following the grading convention in Dutch education system. 
Vollaard had to convert ``$9-$'' (almost worthy of the qualification ``very good'') into a number.  It seems that he rounded it to $9$, but one might just as well have made it $9-\epsilon$ for some choice of $\epsilon$, for instance $0.1$, $0.03$ or $0.01$. We compared the results obtained using various conventions for dealing with the ``broken'' grades, and it turned out that the choice of value of $\epsilon$ had major impact on the statistical significance of the ``just significant'' or ``almost significant'' variables of main interest (supplier and distance). Also, whether one followed standard strategies of model selection based on leaving out insignificant variables has major impact on the significance of the variables which are left in. The ``borderline cases'' can move in either direction.

It is well known that when multicollinearity is present in linear regression analysis, the phenomenon that regression estimates are highly sensitive to small perturbations in model misspecification is commonplace and even to be expected, as noted for instance by \citet{winship2016multicollinearity}. Multicollinearity is here a consequence of \emph{confounding}. The most important factors from a statistical point of view are badly confounded with the factors of most interest. Discretizing continuous variables and grouping categories of discrete variables changes especially the apparent statistical significance of the variables of most interest (since their effects, if they exist at all, are pretty small).
A common way of measuring multicollinearity in a regression model is to compute the condition number of the design matrix, and its value for the second model in Vollaard's first working paper was about 929.  According to usual statistical convention, a value larger than 30 indicates strong multicollinearity.  Consequently, it is hardly surprising that tiny changes to which variables are included and which are not included, as well as tiny changes in the quantification of the variable to be explained, keep changing the statistical significance of the variables which interest us the most.
Furthermore, if one would investigate whether interaction terms for (statistically) highly important variables are needed, this led immediately to singularity of the design matrix, which is again unsurprising given its high condition number. 



In view of the earlier claims of regional bias, we decided to map the outlets and their scores.  This also allowed us to compare the spatial distribution of outlets entering the test over the two years. We also tried to model the effect of ``location''.  As a \emph{toy model} for demonstration, we find that there was certainly enough data to fit a quadratic effect of (latitude, longitude), alongside the already included variables but instead of ``distance from Rotterdam''.  The spatial quadratic terms give us a just significant p-value (F-test), just as the dummy variable for ``distance from Rotterdam'' did.  Fitting the same regression with only linear spatial terms instead of quadratic terms leads to F-test for the two spatial terms together having an impressive p-value of around $0.005$.  It seems to us from this, that distance from Rotterdam, discretized to a binary variable (greater than or less than 30 km), is a poor description of the effects of 2D ``space''. A small distance from the West Coast of the Netherlands along the provinces of South and North Holland leads to high scores. At the Southern and Eastern extremities of the country, scores are a bit lower on average; but also, the spatial density of outlets participating in the test decreases as the distance from the sea increases.

All this is in retrospect hardly surprising. We are talking about a delicacy associated with the summer vacations on the coast of huge numbers of both German and Dutch tourists, as well as with busy markets in the historic towns and cities visited by all foreign tourists, and finally with the high population concentration of a cluster of large cities not far from the same coast. Actually, ``Dutch New Herring'' has been aggressively marketed from the old harbour town of Scheveningen as a major tourist and folkloristic attraction only since the 50s of the last century, in order to help the local tourist industry and the ailing local fishing industry.

The spatial effects we see are exactly what one would expect. However, this is also associated with new entrants to the AD herring test; and new entrants often get poor scores. The effects of space and time are utterly confounded, and any possible bias towards outlets supplied by Atlantic simply cannot be separated from the enormous variation in scores. Recall that we saw in the original simple regression model (after removal of disqualified outlets) typical prediction errors of $\pm 2$ in the mid-range outlets, $\pm 1$ at the extremes. 
New outlets may correspond to enterprising newcomers in the fish restaurant business, hoping to break open some new markets.

Of course, as well as long range effects which might well be describable by a quadratic surface, ``geography'' is very likely to have short range effects. Town and country are different. Rivers and lakes form both trade routes and barriers and have led historically to complex cultural differences across this small country, which modern tourists are unlikely to begin to notice.

The ability to make these plots also allows us to look at exactly \emph{where} the Atlantic supplied outlets are located.  Most are in the vicinity of Rotterdam and The Hague, but an appreciable number are quite far to the East, independently of which of the two later classifications are used. These geographic ``outliers'' tended not to get very high final scores.

\begin{figure}[htp]
\begin{subfigure}{0.48\textwidth}
\includegraphics[width=\textwidth]{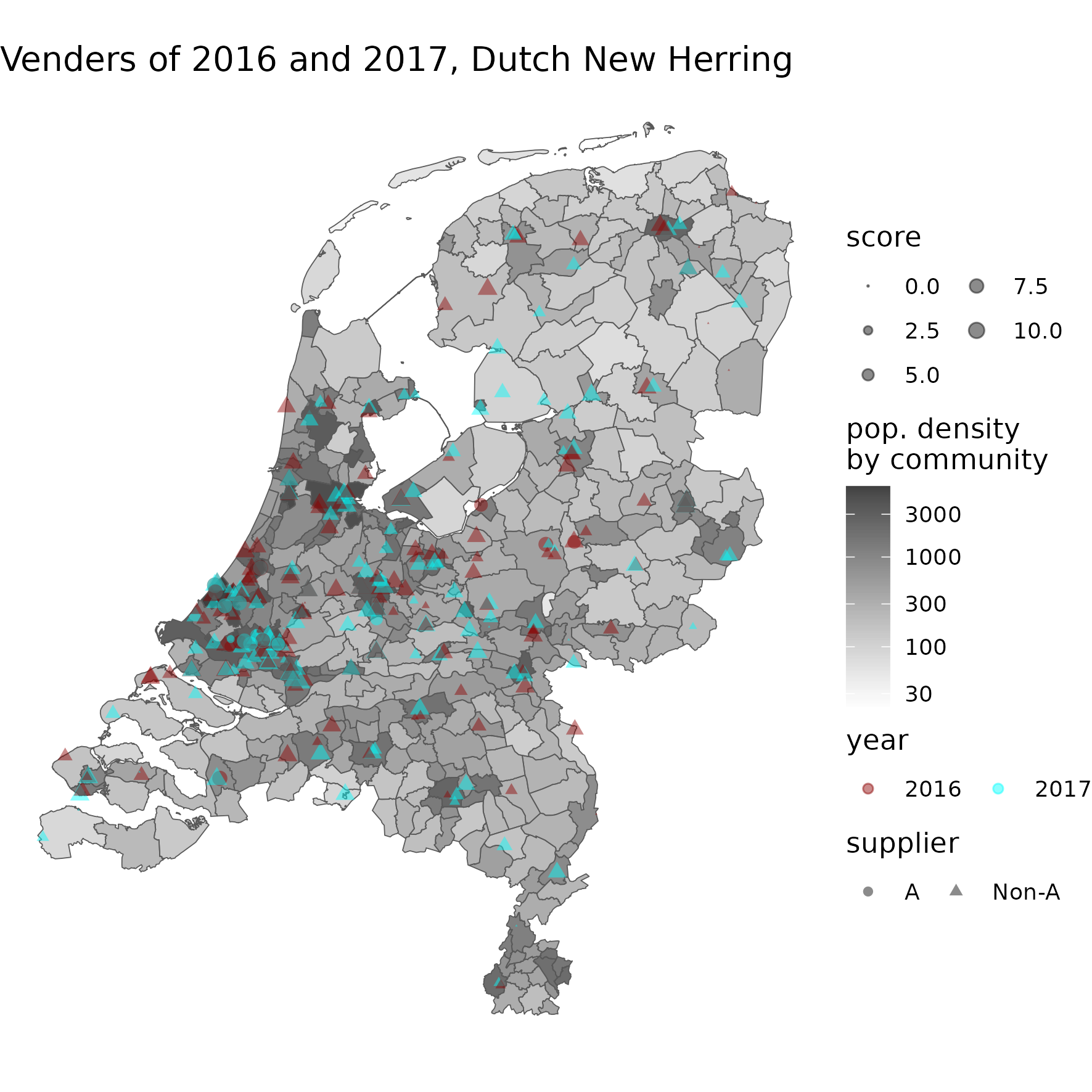}
\end{subfigure}
\begin{subfigure}{0.48\textwidth}
\includegraphics[width=\textwidth]{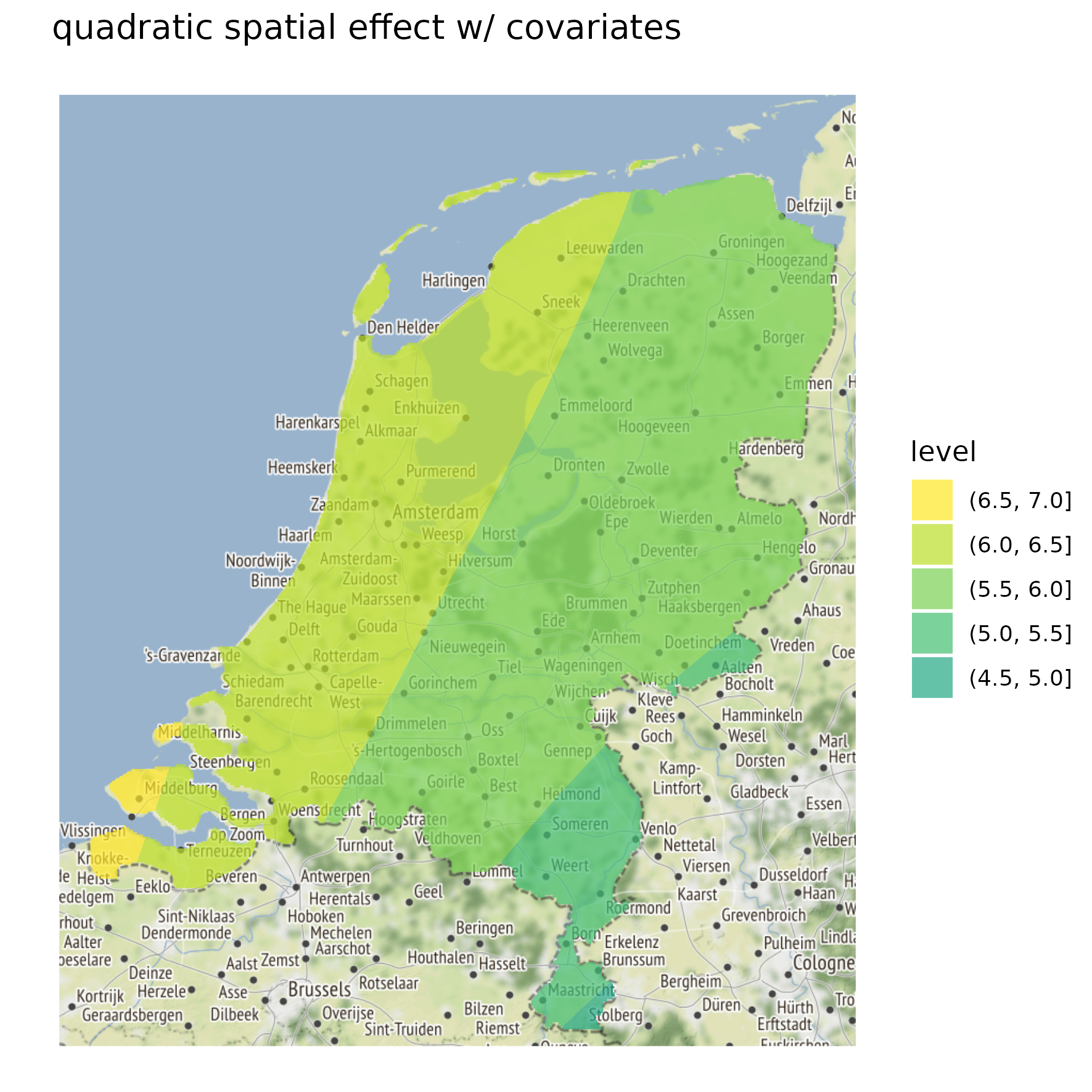}
\end{subfigure}
\caption{Left: new entrants score lower and lie in new distant areas.  The supplier follows the classification of the AD, and the \texttt{A} label indicates Atlantic outlets.  The Netherlands is divided into administrative communities, and the population density of each community of plotted in the background.  Right: spatial effect of large distance from Rotterdam is huge.  The overall level of the surface visualized on the right has been set so that outlets in Utrecht in the centre of the county have a predicted final score of 6.  We used everything from Vollaard's first model (no removal of ``disqualified'' outlets), except for k30.  The p-value obtained from the F-statistics for the quadratic spatial terms is about $0.046$, which is of the same order as $0.022$ --- the p-value for k30 in Vollaard's first working paper \citep*{vollaard1}.}
\label{fig:maps}
\end{figure}

\subsection{New tools from causal inference}

The great advancements in causal inference in the past decades have provided useful new tools for evaluating the goodness of fit of descriptive models including easily accessible R packages, such as \texttt{twang}. 
This uses the key idea of propensity score weighting \citep{mccaffrey2004propensity}, assigning estimated weights to each data point to properly account for the systematic differences in other features between the two groups --- `Atlantic' or not `Atlantic' in our case. 
Such methods estimate the propensity scores using generalized boosted model, which in turn is built upon aggregating simple regression trees.  This essentially nonparametric approach can take account empirically of the problems of how to model the effect of continuous or categorical variables, including their interactions, by a data-driven opportunistic search, validated by resampling techniques.  Being non-parametric, the precision of the estimate of the effect of ``Atlantic'' will be much less than the apparent precision in an arbitrarily chosen parametric model. Our point is, that that precision is probably illusory.

We have performed this analysis, and have documented it in the \texttt{R} scripts in our GitHub repository.  It was a standard and straightforward practice, where we fed the dataset, including all records in both 2016 and 2017, to the \texttt{twang} package with the variables in their original measurements and followed the steps recommended by the package. 
The analysis then proceeded to calculate the so-called Average Treatment Effect on the Treated (ATT) --- a quantity measuring the causal difference resulted in by the \textit{treatment}, i.e., whether the outlet was supplied by Atlantic or not, among \textit{all} outlets.  The results in fact were seemingly supportive of Vollaard's claims where the ATT of being supplied by Atlantic is somewhat statistically significant.  However, this is not the last word on the matter.  Recalling our discussions on the repeat issue, we added the dummy variable indicating whether an outlet had appeared in both 2016 and 2017 to the features, and the said ATT immediately became statistically insignificant,  and became even more insignificant when we excluded the outlets with zero final scores. 
{We remark that such sensitivity of the statistical ``significance'' with respect to slight tweaks of the data and/or the model, and/or even how data is specified, is precisely the evidence we look for in the sensitivity analysis, in both this section and Section~\ref{atlantic}.  It is inadvisable (or even harmful in some cases) to take such a statistical-significance based ``discovery'' seriously and to make further implicit causal claims, as Dr Vollaard has done.}

\subsection{The second argument}

The declared aim of the new paper  \citet{vollaard2021bias} is to show that the outcome of the AD Herring Test ranking was not determined by the evaluations written down by the testers. This is supposed to imply that the results must have been seriously influenced by favouritism toward outlets supplied by \emph{Atlantic}: the favoured outlets have been engineered to come out in the top ten. Interestingly, the outlets attributed to \emph{Atlantic} have changed again. There are now 27 of them, and they differ on six outlets from the list obtained for us by AD from Atlantic.\footnote{We approached Vollaard and van Ours to discuss the differences, but they decline communication with us.}

Vollaard and van Ours make what they call the ``crucial identifying assumption'' that \emph{the reviewers’ assessment of quality is} [they mean: should be] \emph{fully reflected in the published ratings and verbal judgements of individual attributes}. Certainly it is crucial to their whole paper but is it justified?
We are not aware of any claim made by AD that their ranking was based \emph{only} on the features concerning the taste of the herring about which they explicitly commented and also assigned scores to. When one evaluates restaurants, one is also) interested in the friendliness of the waiters, the helpfulness of the staff, the cleanliness and attractiveness of the establishment, the price. (This is also reflected in the two verbal assessments of each outlet; the one written immediately after the visit, and a final one written after the laboratory outcomes come in.)  The AD Herring Test rightfully evaluated the herring eating experience, on site, with an aim to ranking the sites in order to advise consumers where to go and where not to go. 

But even if we accept this ``identifying assumption'', Vollaard and van Ours need to make the further assumption that when they predict the score using a particular regression model, their particular model \emph{can} fully reflect the published ratings and \emph{does} take account of all the information written down by the tasting team about their experience. However, they still make some model choices based on statistical criteria, and this essentially comes down to reducing variance by accepting increased bias. A more parsimonious model can be better overall at prediction.  But this does not mean that it does \emph{accurately capture everything} expressed by the tester's written down evaluations. That aim is a \emph{fata morgana}, an illusion.

An interesting change of tactic is that instead of modelling the ``final score'', they now model  the ``provisional score'' also available from the AD website, which was written down by the testers at the point of sale, before knowing the ``objective variables'' temperature, microbiological contamination, weight and price per 100 g. Apart from this, something like Vollaard's original model was run again. The dependent variable was the provisional rating; the explanatory variables were ripening, quality of cleaning, and cleaned on site, together with a new quality variable which they came up with by themselves. The AD website contains, for each outlet, a one- or two-sentence verbal description of the jury's experience. It also contains another very verbal final summary, but they leave this out, since their plan is only to study \emph{recorded actual taste impressions obtained at point of sale}. We know that the provisional score did include the half point reduction when the herring was not cleaned on site, so that reduction effectively has to be ``undone'' by including just that single ``non-taste'' variable.

In any case, they need to quantify the just mentioned verbal evaluation of taste. As they revealingly say, the sample is too small to use methods from Artificial Intelligence to take care of this.  Instead, Vollaard and van Ours construct a new six-category variable themselves ``by hand'', with categories \emph{disgusting, unpleasant, bland, okay, good, excellent}. They ``subjectively'' allocated one of these six "Overall quality" evaluations to each of the outlets, using the recorded on-site verbal evaluation of the panel. They obtained four ``replicates'' by having four persons each independently figure out their own judgement under the same classification scheme, while also only given the verbal descriptions, and some explanation of what to look for: four sensory dimensions of eating a herring: taste, smell, appearance (both interior and exterior), texture. The category ``bland'' seems to account for the evaluation ``green'' of ripening, discussed before, and grouped with ``lightly matured''. The subsequent results do not appreciably differ when they replace their score with any of their four replicates.

In the subsequent statistical analysis there was still no correction for heteroscedasticity, no sign of inspection of diagnostic plots based on residuals, no correction for correlation of the error term for return participants. The data (their new score plus four replicate measurements of it) can be found on van Ours' website, so we were able to replicate and add to their statistical analyses.
We discovered the same serious heteroscedasticity --- shrinking variance near the endpoints of the scale. We found the same problem that small changes to model specification, and especially addition of more subtle modelling of location, caused severe instability of the just significant effect of  ``Atlantic''. The model fit is overall somewhat improved, $R^2$ is now a bit above 90\%; the problem with ``disqualified'' outlets has become smaller. Since the variables which were not known during the initial visit are not included, the model has less explanatory variables, and the authors could perform some tests of goodness of fit (e.g., investigation of linearity) in various ways. 

They also performed a goodness of fit procedure somewhat analogous to our modern approach using propensity score matching. They searched for a large group of outlets very close in all their taste variables and moreover including a sizeable proportion of Atlantic outlets. This led to a group of about thirty ``high achievers'' including many Atlantic outlets and mainly in the region of The Hague and Rotterdam.  The difference between the average ``provisional score''  of Atlantic and non-Atlantic outlets in this \emph{particular} small group, uncorrected for covariates, was a significant approximately 0.5.  We point out that as the group is made smaller and yet more homogenous, bias might decrease but variance will increase, statistical significance at the 1\% level will not be stable.  It would not surprise us if a few misidentified outlets and slightly modified ``verbal judgements'' could ruin their conclusion, even at the magic 5\% level. 

To sum it up, Vollaard and van Ours claim to have found a consistently just significant effect of ``Atlantic'' of about a third of a point. They claim it is rather stable under variations of their model. We found it to be very unstable, especially when we model ``location'' in a more subtle way. They also stated ``\emph{given our exclusive focus on how the overall rating is compiled, our results may only reflect a part of the bias that results from the conflict of interest}''. 
In our opinion, even if there is a possibly small systematic advantage of Atlantic outlets (which may be attributed to all kinds of fair reasons), it is irresponsible to claim that it can \emph{only} be due to favouritism and that it \emph{must} be an underestimate.  We see plenty of evidence that the effect is due to misspecification and the effects of time and space.  

\section{Something Ends, Something Begins}\label{whatnext}


As it probably should be, when the AD Herring Test came to its end, another newspaper \emph{Leiden Courant} stepped in and started its own test.  It was designed to avoid all suggestions of bias and favouritism. The organizers were advised by experts in the field of consumer surveys. Fish was collected from participating outlets and brought, well refrigerated, to a central location, where each of a panel of tasters got to taste the herring from all participating outlets, without knowing the source, not unsimilar to the double-blind practice in drug trials. Initially, numbers of participants were quite small. The panel consisted of 15 celebrities and over the years has included TV personalities, scientists, writers, football players, \dots, even the Japanese Ambassador. Each year a brand-new panel is put together. Within a few years, however, the test started to run into problems. Outlets were not keen to come back and be tested again, the test had to be expanded from the regional to the national level, but never achieved the kind of fame which the AD herring test had ``enjoyed''. At some point it was abandoned by the newspaper but rebooted a second time by an organization specializing in promotions and public events.  The test is called the National Herring Taste Competition.  No ``scientific'' measurements are taken of weight, fat content, temperature or whatever: the panel is meant to go purely on taste.

Certainly this new style of testing appeals to today's public who probably do not have much respect for ``experts''. One can wonder how relevant its results are to the consumer. How a delicacy tastes does depend on the \emph{setting} where it is consumed. It also depends on the temperature at which it is served and yet in this test, temperature is equalized. In the new herring test, the setting is an expensive restaurant in a beautiful location and the fellow diners are famous people. The \emph{real life experience} of Dutch New Herring is influenced by the ritualistic delight of seeing how the fish is personally prepared for you. As we mentioned above, it is maybe good to know which supermarkets have the best herring in the freezer, but it is not clear that this question needs to be answered anew every year with fanfares and an orchestra.

\section{Conclusions}\label{conclusion}

The hypothesis that the AD Herring Test was severely biased is not supported by the data. Obviously, a conflict of interest was present, and that was not good for the reputation of the test. The test probably died because of its own success, growing from some fun in a local newspaper to a high profile national icon. Times have changed, people do not have such trust in ``experts'' as they used to, and everyone knows that they themselves know what a good Dutch herring should taste like. The Herring Test did successfully raise standards and not surprisingly, superb Dutch New Herring can now be enjoyed at many more outlets than ever before.

Vollaard's analyses have some descriptive value, but with only a little more work, he could have discovered that his model was badly wrong, and in as much as the final ranking can be predicted from the measured characteristics of the product, much more sophisticated modelling is needed.  The aspects of space and time deserve further investigation, and it is a pity that his immature findings caused the AD Herring test to be so abruptly discontinued. The present organizers of the rebooted New Herring Taste Test might want to bring back some of the ``exact measurements'' of the AD Test, and new analyses of data from a new ``stable'' annual test would be interesting.

In this case Ben Vollaard seems to us to have been a victim of the currently huge pressure on academics to generate media attention by publishing on issues of current public interest. This leads to immature work being fed to the media without sufficient peer review in terms of discussion by the relevant community of experts through seminars, dissemination of preprints, and so on. Sharing of data and of data analysis scripts should have started right at the beginning.



\section{Conflict of interest}\label{COI}

The second author was paid by a well known law firm for a statistical report on Vollaard's analyses. His report, dated April 5, 2018, formed the basis of earlier versions of this paper. He also reveals that the best Dutch New Herring he ever ate was at one of the retail outlets of \emph{Simonis} in Scheveningen. They got their herring from the wholesaler \emph{Atlantic}. He had this experience before any involvement in the Dutch New Herring controversies, topic of this paper. 

\bibliography{./herring.bib%
}

\end{document}